\documentclass[12pt]{article}

\usepackage{amssymb}
\usepackage{amsmath}

\usepackage{epsfig}

\begin{document}%

\title{\bf RS model with a small curvature and two-photon production
at the LHC}

\author{A.V. Kisselev\thanks{Electronic address:
alexandre.kisselev@ihep.ru} \\
\small Institute for High Energy Physics, 142281 Protvino, Russia}

\date{}

\maketitle

\thispagestyle{empty}

\bigskip


\begin{abstract}
The $p_{\perp}$-distribution for the diphoton production at the
LHC is calculated in the modified Randall-Sundrum model with a
small curvature of the space-time $\kappa$, and 5-dimensional
Planck scale $M_5$ in the TeV region. The discovery limits on
$M_5$ are obtained to be 9.4 TeV and 11.6 TeV for the integrated
luminosities 30 fb$^{-1}$ and 100 fb$^{-1}$, respectively. These
limits do not depend on $\kappa$ provided it is much smaller than
$M_5$. The account of graviton widths is a crucial points of our
calculations.
\end{abstract}



\section{RS1 model with a small curvature}
\label{sec:RS}

In this Section we will consider in details the Randall-Sundrum
(RS) model with the small curvature. The main features and
differences from a conventional (large curvature) scenario will be
pointed to.

The RS model~\cite{Randall:99} is a theory with one extra
dimension (ED) in a slice of the AdS$_5$ space-time. It has the
following background warped metric:
\begin{equation}\label{metric}
ds^2 = e^{2 \kappa (\pi r_c - |y|)} \, \eta_{\mu \nu} \, dx^{\mu}
\, dx^{\nu} + dy^2 \;,
\end{equation}
where $y = r_c \theta$ ($-\pi \leqslant \theta \leqslant \pi$),
$r_c$ being the ``radius'' of the ED, and  $\eta_{\mu \nu}$ is the
Minkowski metric. The points $(x_{\mu}, y)$ and $(x_{\mu}, -y)$
are identified, so one gets the orbifold
$\mathrm{S}^1/\mathrm{Z_2}$.

The parameter $\kappa$ defines a 5-dimensional scalar curvature of
the AdS$_5$ space. For the sake of simplicity we will call
$\kappa$ ``curvature''.

The so-called RS1 model has two 3D branes with equal and opposite
tensions located at the points $y = \pi r_c$ (called the TeV
brane) and  $y = 0$ (referred to as the Plank brane). If $k
> 0$, the tension on the TeV brane is negative, whereas the
tension on the Planck brane is positive. All the SM fields are
constrained to the TeV brane, while the gravity propagates in four
spacial dimensions.

It is necessary to note that metric \eqref{metric} is chosen in
such a way that 4-dimensional coordinates $x_{\mu}$ are Galilean
on the TeV brane where all the SM field live, since the warp
factor is equal to unity at $y = \pi r_c$.

By integrating 5-dimensional action over variable $y$, one gets an
effective 4-dimensional action, that results in the ``hierarchy
relation'' between the reduced Planck scale
$\bar{M}_{\mathrm{Pl}}$ and 5-dimensional reduced gravity scale
$\bar{M}_5$:
\begin{equation}\label{RS_hierarchy_relation}
\bar{M}_{\mathrm{Pl}}^2 = \frac{\bar{M}_5^3}{\kappa} \left( e^{2
\pi \kappa r_c} - 1 \right) \;.
\end{equation}

The reduced gravity scale $\bar{M}_5$ is related to the Planck
mass $M_5$ by the equation
\begin{equation}\label{reduced_scale}
M_5 = (2\pi)^{1/3} \, \bar{M}_5 \simeq 1.84 \, \bar{M}_5 \;.
\end{equation}
Unless explicitly stated otherwise, we consider the \emph{reduced}
5-dimensional Planck scale.

From the point of view of a 4-dimensional observer located on the
TeV brane, there exists an infinite number of graviton KK
excitations with the masses
\begin{equation}\label{graviton_masses}
m_n = x_n \, \kappa, \qquad n=1,2 \ldots \;,
\end{equation}
where $x_n$ are zeros of the Bessel function $J_1(x)$. Note that
$x_n \simeq \pi (n + 1/4)$.

The interaction of the KK gravitons with the the SM fields
 on the TeV brane is described by the Lagrangian:
\begin{equation}\label{Lagrangian}
\mathcal{L}_{int} = - \frac{1}{\bar{M}_{\mathrm{Pl}}} \, T^{\mu
\nu} \, G^{(0)}_{\mu \nu} - \frac{1}{\Lambda_{\pi}} \, T^{\mu \nu}
\, \sum_{n=1}^{\infty} G^{(n)}_{\mu \nu} \;.
\end{equation}
Here $T^{\mu \nu}$ is the energy-momentum tensor of the matter,
and $G^{(n)}_{\mu \nu}$ is the graviton field with the KK-number
$n$. The parameter $\Lambda_{\pi}$,
\begin{equation}\label{lambda}
\Lambda_{\pi} = \bar{M}_5 \,\left( \frac{\bar{M}_5}{\kappa}
\right)^{\! 1/2} ,
\end{equation}
is the physical scale on the TeV brane.

Note that in most of the papers which treated the RS model
(including Ref.~\cite{Randall:99}), the background metric was
taken to be
\begin{equation}\label{metric_uncorr}
ds^2 = e^{-2\kappa|y|} \, \eta_{\mu \nu} \, dx^{\mu} \, dx^{\nu} +
dy^2 \;,
\end{equation}
instead of expression \eqref{metric}. In such a case, the
hierarchy relation looks like
\begin{equation}\label{RS_hierarchy_relation_uncorr}
\bar{M}_{\mathrm{Pl}}^2 = \frac{\bar{M}_5^3}{\kappa} \left( 1 -
e^{-2 \pi \kappa r_c} \right) \;,
\end{equation}
with the graviton masses
\begin{equation}\label{graviton_masses_uncorr}
m_n = x_n \, \kappa \, e^{- \pi \kappa r_c}, \qquad n=1,2 \ldots
\;.
\end{equation}

In order the hierarchy relation
\eqref{RS_hierarchy_relation_uncorr} to be satisfied, while the
lightest graviton masses to be around one TeV, one has to
introduce two huge mass scales (\emph{large curvature option}),
\begin{equation}\label{scale_relation_uncorr}
\kappa \sim \bar{M}_5 \sim \bar{M}_{\mathrm{Pl}} \;,
\end{equation}
and take $\kappa r_c \simeq 12$.

Thus, one obtains a series of graviton resonances with the
lightest KK mode around one TeV. The experimental signature of the
the RS model with the large curvature
\eqref{scale_relation_uncorr} is a real or virtual production of
the massive KK graviton resonances.

There exists a serious shortcoming of the scenario with the
curvature and fundamental gravity scale being of the order of the
Planck mass~\eqref{scale_relation_uncorr}. Namely, \emph{kinetic
terms} of the graviton fields on the TeV brane \emph{do not have a
canonical form}, and Lorentz indices are raised with the Minkowski
tensor, while the metric in the coordinates $x_{\mu}$ is $e^{- \pi
\kappa r}\eta_{\mu \nu}$ (for details, see Ref.~\cite{Boos:02}).

The correct interpretation of the effective 4-dimensional theory
can be achieved by changing variables:
\begin{equation}\label{change_variables}
x^{\mu} \rightarrow z^{\mu} =  x^{\mu} e^{- \pi \kappa r_c} \;.
\end{equation}
As one can see, metric \eqref{metric_uncorr} turns into metric
\eqref{metric} under such a replacement.

In the present paper we will use an approach based on the
metric~\eqref{metric} (\emph{small curvature
option}~\cite{Giudice:05}-\cite{Kisselev:06}) and put
\begin{equation}\label{scale_relation}
\bar{M}_5 = (1 \div 20)  \mathrm{\ TeV} \;, \qquad \kappa = (0.1
\div 1) \mathrm{\ GeV}  \;.
\end{equation}
In order the hierarchy relation \eqref{RS_hierarchy_relation} to
be satisfied, it is then necessary to take $r_c \kappa \simeq 8
\div 9.5$, i.e. $r_c \simeq 0.15 \div 1.8 \mathrm{\ fm}$. Thus, no
large scales are introduced, contrary to the large curvature case.
In what follows, the RS1 model with the small curvature will be
referred to as \emph{large warped dimension} scheme.

Instead of the fact that the size of the ED $r_c$ is of the order
of 1 fm, and masses of the first KK gravitons are relatively
small, they give a negligible correction to the Newton law.
Indeed, the Newton potential between two test masses $m_1$ and
$m_2$ separated by the distance $r$ is equal to
\begin{align}\label{grav_potential}
V(r) &= G_N \, \frac{m_1 m_2}{r} + \int\limits_{m_1}^{\infty} \!
\frac{dm}{\pi \kappa} \, G_N \, \frac{m_1 m_2}{r} e^{-m r}
\nonumber
\\
&= G_N \, \frac{m_1 m_2}{r} \left( 1 + \frac{e^{-m_1 r}}{\pi
\kappa r}\right) \;.
\end{align}
Since $m_1 = x_1 \kappa$, where $x_1 = 3.84$ is the first zero of
the Bessel function $J_1(x)$, we find that the relative correction
from the KK gravitons to the Newton law is less than $2 \cdot
10^{-15}$.

In our scheme, a coupling of the KK graviton with the SM fields
is rather weak,%
\footnote{In the conventional scheme of the RS model,
$\Lambda_{\pi} \simeq 1$ TeV.}
since
\begin{equation}\label{lambda_enum}
\Lambda_{\pi} = 100 \left( \frac{M_5}{\mathrm{TeV}} \right)^{3/2}
\left( \frac{\mathrm{100 \ MeV}}{\kappa} \right)^{1/2} \!
\mathrm{TeV} \;.
\end{equation}
However, in physical matrix elements the smallness of the coupling
$\Lambda_{\pi}^{-2}$ is compensated by the large number of real
gravitons which can be produced or by infinite number of virtual
gravitons. As a result, the matrix elements are defined by the
5-dimensional gravity scale $\bar{M}_5$, not by $\Lambda_{\pi}$ or
$\kappa$ separately~\cite{Kisselev:05,Kisselev:06}. This
circumstance reminds that in the ADD model with flat
EDs~\cite{Arkani-Hamed:98}.

Previously, we studied gravity effects in the RS1 model with the
small curvature in the scattering of ultrahigh neutrinos off the
nucleon~\cite{Kisselev:05}, in exclusive double diffractive events
at the LHC~\cite{Kisselev:diff}, as well as in $e^+e^-$
annihilation into lepton pairs at the ILC~\cite{Kisselev:ILC}.

The goal of the present paper is to estimate gravity effects in a
two-photon production at the LHC. The search limits on the
fundamental Planck scale will be derived which are insensitive to
the curvature of the warped space-time of the model.

\section{Virtual graviton contribution to a diphoton production}
\label{sec:virtual_gravitons}

Diphoton final states are a signature of many interesting physics
processes. For instance, one of the main discovery channels for
the Higgs boson search at the LHC is the $\gamma \gamma$ final
state. An excess of $\gamma \gamma$ production could be a
signature of interactions beyond the SM. In addition, the diphoton
final state is interesting in its own right. Using good energy
resolution of the electromagnetic calorimeter~\cite{CMS:TDR}, the
transverse momentum of the photons, $p_{\perp}$, can be directly
determined with a good precision. A possible excess in
$p_{\perp}$-distribution could indicate effects coming from a new
physics, in particular, from large EDs.

What is the reason to consider large warped dimension scenario?
First, the spectrum of the KK gravitons \eqref{graviton_masses} is
very similar to that in the model with \emph{one}
ED~\cite{Arkani-Hamed:98} (see previous Section). Second, all
matrix elements for the scattering of the SM fields can be
formally obtained from corresponding matrix elements calculated in
the model with one \emph{flat} dimension by using the following
replacement~\cite{Kisselev:05,Kisselev:ILC}:
\begin{equation}\label{RS_vs_ADD}
\bar{M}_{4+1} \rightarrow (2 \pi)^{-1/3} \, \bar{M}_5 \;, \qquad
R_c \rightarrow (\pi \kappa)^{-1} \;.
\end{equation}
Here $\bar{M}_{4+1}$ is a 5-dimensional reduced Planck scale,
$R_c$ being the radius of the extra flat dimension. As a result,
all cross sections appear to be rather large (as in the ADD model
with $4+1$ dimensions). Finally, as was shown in
Ref.~\cite{Kisselev:06}, astrophysical restrictions are not
applied to the RS model with the large ED (i.e. small graviton
masses), if the curvature lies within the limits
mentioned above \eqref{scale_relation}.%
\footnote{Actually, much smaller values of $\kappa$ are
allowed~\cite{Kisselev:06}.}

Let us consider the two-photon production with hight transverse
momenta ($p_{\perp} \ll \sqrt{s}$ is assumed):
\begin{equation}\label{process}
p \, p \rightarrow \gamma \, \gamma + X \;,
\end{equation}
where $X$ denotes a remnant of the colliding protons. The
differential cross section is equal to
\begin{align}\label{cross_sec}
\frac{d \sigma}{d p_{\perp}^2}(p p \rightarrow \gamma \gamma + X)
&= \sum\limits_{a,b} \int\limits_0^1 \! dx_a f_{a/p}(\mu^2, x_a)
\!\! \int\limits_0^1 \! dx_b
f_{b/p}(\mu^2, x_b) \nonumber \\
&\times  \theta(x_a x_b - x_{\perp}^2) \, \frac{\sqrt{x_a
x_b}}{\sqrt{x_a x_b - x_{\perp}^2}} \, \frac{d
\hat{\sigma}}{d\hat{t}}(a b \rightarrow \gamma \gamma) \;.
\end{align}
Here $ f_{a/p}(\mu^2, x_a)$ is the distribution of the parton of
the type $a$ in momentum fraction $x_a$ inside the proton taken at
the factorization scale $\mu$ (this scale will be fixed below).
$d\hat{\sigma}/d\hat{t}$ denotes the cross section of the hard
sub-process $a b \rightarrow \gamma \gamma$. We have introduced
the dimensional variable
\begin{equation}\label{x_trans}
x_{\perp} = \frac{2 p_{\perp}}{\sqrt{s}} \;.
\end{equation}
Throughout the paper, $\hat{s}$, $\hat{t}$ and $\hat{u}$ denote
Mandelstam variables of the \emph{partonic sub-process} ($\hat{s}
+ \hat{t} + \hat{u} = 0$, $\hat{s} = s\,x_a x_b$, $\hat{t}
\hat{u}/\hat{s} = p_{\perp}^2$).

The contribution of the virtual gravitons to the process
\eqref{process} comes from the quark-antiquark annihilation,
\begin{align}
q \, \bar{q} &\rightarrow G^{(n)} \rightarrow \gamma \, \gamma \;,
\label{quark_annihilation}
\intertext{and gluon-gluon fusion,}
g \, g &\rightarrow G^{(n)} \rightarrow \gamma \, \gamma \;.
\label{gluon_fusion}
\end{align}
The matrix elements for both the partonic
sub-processes~\eqref{quark_annihilation}-\eqref{gluon_fusion} are
given by the expression
\begin{equation}\label{matrix_element}
\mathcal{M} = \mathcal{A} \, \mathcal{S} \; .
\end{equation}
The fist factor in the r.h.s. of Eq.~\eqref{matrix_element} is
\begin{equation}\label{tensor_part}
\mathcal{A} = T_{\mu \nu}^{q(g)} \, T^{\gamma \, \mu \nu} -
\frac{1}{3} \, {(T^{q(g)})}^{\mu}_{\mu} \,
{(T^\gamma)}^{\nu}_{\nu} \;,
\end{equation}
where $T_{\mu \nu}^{q(g)}$ is the energy-momentum tensor of the
quark (gluon) field, $T_{\mu \nu}^{\gamma}$ is the photon
energy-momentum tensor.

The graviton exchange in the $s$-channel leads to the following
expression:
\begin{equation}\label{KK_sum}
\mathcal{S}(\hat{s}) = \frac{1}{\Lambda_{\pi}^2}
\sum_{n=1}^{\infty} \frac{1}{\hat{s} - m_n^2 + i \, m_n \Gamma_n}
\;.
\end{equation}
Here $\Gamma_n$ denotes a total width of the graviton with the KK
number $n$ and mass $m_n$:
\begin{equation}\label{graviton_widths}
\Gamma_n = \eta \, m_n \left( \frac{m_n}{\Lambda_{\pi}} \right)^2
\;,
\end{equation}
with $\eta \simeq 0.09$~\cite{Kisselev:05_2}. The width is small
provided $n$ is not extremely large.

Then the relevant partonic cross sections are (see, for instance,
formulae in Appendix of Ref.~\cite{Giudice:05}):
\begin{align}
\frac{d\hat{\sigma}}{d\hat{t}}(q \bar{q} \rightarrow \gamma
\gamma) &= \frac{\hat{t}^2 + \hat{u}^2}{192 \pi
\hat{s}^2\hat{t}\hat{u}}
\left| 2e_q^2 - \hat{t} \hat{u} \, \mathcal{S}(\hat{s}) \right|^2 \;,
\label{quark_cross_sec} \\
\frac{d\hat{\sigma}}{d\hat{t}}(gg \rightarrow \gamma \gamma) &=
\frac{\hat{t}^4 + \hat{u}^4}{512 \pi \hat{s}^2} \left|
{S}(\hat{s}) \right|^2 \;, \label{gluon_cross_sec}
\end{align}
where $e_q$ is the electric charge of the quark $q$.

In the region $\sqrt{\hat{s}} \sim \bar{M}_5 \gg \kappa$, the sum
\eqref{KK_sum} can be calculated analytically, an explicit form of
the function $\mathcal{S}(\hat{s})$ was derived in
Ref.~\cite{Kisselev:06}:
\begin{equation}\label{grav_propagator}
\mathcal{S}(\hat{s}) = - \frac{1}{4 \bar{M}_5^3 \sqrt{\hat{s}}} \;
\frac{\sin 2A + i \sinh 2\varepsilon }{\cos^2 \! A + \sinh^2 \!
\varepsilon } \;,
\end{equation}
where
\begin{align}
A &= \frac{\sqrt{\hat{s}}}{\kappa} \;,
\label{A} \\
\varepsilon &= \frac{\eta}{2} \Big(
\frac{\sqrt{\hat{s}}}{\bar{M}_5} \Big)^3 \;.
\label{epsilon}
\end{align}

Should we ignore the widths of the massive gravitons, and replace
the summation in KK number \eqref{KK_sum} by the integration in
graviton masses using the relation $dn = dm/(\pi \kappa)$, we get
\begin{equation}\label{KK_sum_zero_widths}
\mathrm{Im} \,\mathcal{S}(\hat{s}) = - \frac{1}{2 \bar{M}_5^3
\sqrt{\hat{s}}} \;, \qquad  \mathrm{Re} \, \mathcal{S}(\hat{s}) =
0 \;,
\end{equation}
in contrast to formula \eqref{grav_propagator}.

At $\sqrt{\hat{s}} \gtrsim 3.5 \, \bar{M}_5$, we get from
\eqref{grav_propagator}-\eqref{epsilon}:
\begin{equation}\label{sum_zero_widths_app}
\mathrm{Im} \,\mathcal{S}(\hat{s}) \simeq - \frac{1}{2 \bar{M}_5^3
\sqrt{\hat{s}}} \;, \qquad  \mathrm{Re} \, \mathcal{S}(\hat{s}) <
0.05 \, \mathrm{Im} \,\mathcal{S}(\hat{s}) \;.
\end{equation}
The inequality $\sqrt{\hat{s}} > 3.5 \, \bar{M}_5$ is equivalent
to the inequality $\Delta m_{KK} < \Gamma_n$, where $\Delta
m_{KK}$ is the mass splitting, and $\Gamma_n$ is the graviton
width for relevant KK numbers (corresponding to $m_n \sim
\sqrt{\hat{s}})$~\cite{Kisselev:06}. In such a case, one may
regard a set of narrow graviton resonances to be a continuous mass
spectrum.

However, the kinematical region for our treatment is
$\sqrt{\hat{s}} \leq \sqrt{s} < 3 \bar{M}_5$. In this case,
expressions \eqref{KK_sum_zero_widths} become incorrect, and
formula \eqref{grav_propagator}, which takes into account the
\emph{nonzero widths} of the KK gravitons, should be used.

In Appendix~A we will demonstrate that our formula
\eqref{grav_propagator} is a correct expression for the function
$\mathcal{S}(s)$ \eqref{KK_sum}.

\section{5-dimensional Planck scale: LHC search limits}
\label{sec:LHC_limit}

The main goal of this Section is to obtain the LHC search limit
for the 5-dimensional Planck scale $\bar{M}_5$.

Recently, the RS1 model with the large extra dimension has been
checked by the DELPHI Collaboration~\cite{LEP_limit}. The gravity
effects were searched for by studying photon energy spectrum in
the process $e^+e^- \rightarrow \gamma + E_{\perp}\hspace{-6mm}
\diagup \hspace{2mm}$. No deviations from the SM prediction were
seen. The limit on $M_5$ obtained is 1.69 TeV at $95\%$
CL~\cite{LEP_limit}. It corresponds to the following bound on the
\emph{reduced} fundamental scale (see the relation between $M_5$
and $\bar{M}_5$ \eqref{reduced_scale}):
\begin{equation}\label{LEP_limit}
\bar{M}_5 > 0.92 \mathrm{\ TeV}\;.
\end{equation}

The search for large EDs (with flat metric) in the diphoton
channel using of $\approx 200 \mathrm{\ pb}^{-1}$ of data
collected by the CDF and D\O~experiments at $\sqrt{s} = 1.96$ TeV
(Run II) have been presented in Refs.~\cite{Tevatron_ADD}. The
$p_{\perp}$-distribution up to $\sim 200$ GeV has been measured.
The data are in a good agreement with the SM background.
D\O~Collaboration has also performed the search for the massive
gravitons (warped metric with the large curvature) in the diphoton
channel using high integrated luminosity~\cite{Tevatron_RS}. No
evidence for resonant production of the gravitons has been found.

We should also mention the preliminary analysis by the CDF
Collaboration based on $1.2 \mathrm{\ fb}^{-1}$ of
data~\cite{CDF_preliminary}. No significant excess of the data
over the expected background in $\gamma \gamma + X$ events was
observed in $m(\gamma \gamma)$ and $p_{\perp}$-distribution.

\begin{figure}[htb]
\begin{center}
\epsfysize=8cm \epsffile{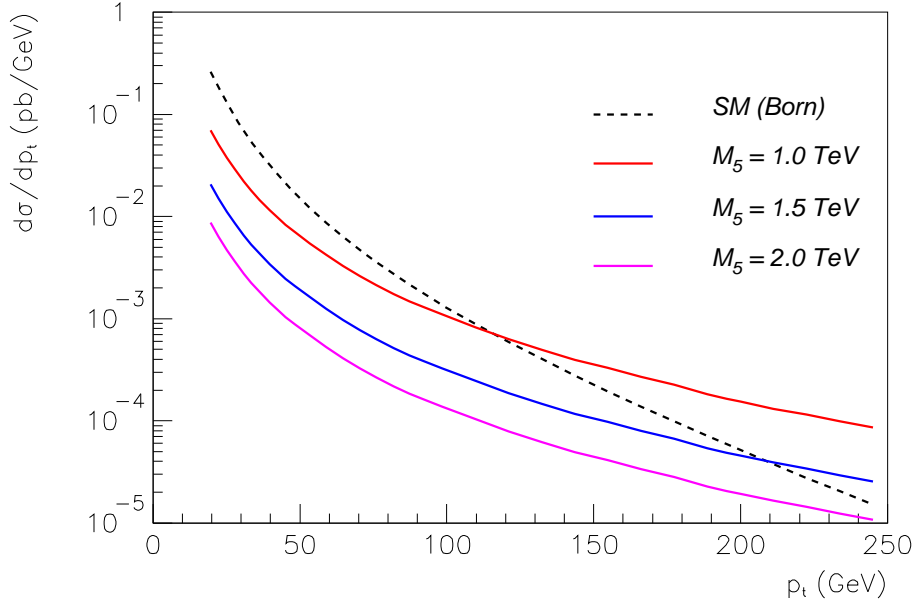}
\end{center}
\caption{The graviton contributions (including the interference
term) to the process $p \, \bar{p} \rightarrow \gamma \, \gamma +
X$ (solid curves) vs. SM contribution (dashed curve) at the
Tevatron. In this and all subsequent figures, $M_5$ denotes the
\emph{reduced} fundamental Planck mass.}
\label{fig:Tevatron_SM_grav}
\end{figure}

In Figs.~\ref{fig:Tevatron_SM_grav} we present the result of our
calculations of gravity effects in the diphoton production $p
\bar{p} \rightarrow \gamma \gamma + X$ at the Tevatron. We used a
set of parton distribution functions (PDFs) from
Ref.~\cite{Alekhin:05} based on an analysis of charged-leptons
proton/deutron data on deep inelastic scattering collected in the
SLAC-CERN-HERA experiments.

Generally, \emph{both} PDFs \emph{and} differential cross sections
in \eqref{cross_sec} should depend on the factorization scale
$\mu$ due to higher order corrections, that results in
$\mu$-independent cross section of the diphoton
production~\eqref{process}. We restrict ourselves by first order
expressions for the partonic cross sections
\eqref{quark_cross_sec}-\eqref{gluon_cross_sec}, and the
factorization parameter is taken to be equal to the relevant mass
scale $\mu = \sqrt{\hat{s}}$. A possible $\mu$-dependence will be
analyzed below (see our comments after
Eq.~\eqref{search_limit_reduced}).

Let $N_S$ be a number of signal events, $N_B$ - number of
background events. We define the statistical significance
$\mathbb{S} = N_S/\sqrt{N_B}$, and require a $5 \sigma$ effect.
Then we obtain the following bound from the Tevatron
data%
\footnote{Let us notice, this bound is not the main goal of the
paper.}
(with the integrated luminosity $\mathcal{L} = 1 \mathrm{\
fb}^{-1}$):
\begin{equation}\label{Tevatron_limit}
\bar{M}_5 > 0.81 \mathrm{\ TeV}\;.
\end{equation}
In calculating numbers of events, we used a K-factor 1.5 for the
SM background, while a conservative value of K=1 was taken for the
signal.

Let us stress that in our approach, the limit on $\bar{M}_5$ does
not depend on the value of the parameter $\kappa$, provided the
inequality $\kappa \ll \sqrt{s}$ is satisfied (see
Section~\ref{sec:details} for details).

We expect that the gravity effects related with the virtual gluon
exchange will be more significant at the LHC, since
\begin{equation}\label{SM_dependence}
\frac{d \sigma(\mathrm{SM})}{d p_{\perp}} = \frac{1}{s^{3/2}}
f(x_{\perp}, \ln s) \;,
\end{equation}
where $f(x_{\perp}, \ln s)$ depends weakly on $s$ via scaling
violation in PDFs. The gravity term $d \sigma(\mathrm{grav})/d
p_{\perp}$ depends rather slowly on $s$ (see
Eq.~\eqref{M5_dependence}). Thus, we obtain:
\begin{equation}\label{grav_vs_SM}
\frac{d \sigma(\mathrm{grav})}{d \sigma(\mathrm{SM})} \sim \left(
\frac{\sqrt{s}}{\bar{M}_5} \right)^3 \;.
\end{equation}
Correspondingly, the search limit for the LHC can be roughly
estimated to be $\bar{M}_5 = (6 \div 7)$ TeV.

\begin{figure}[htb]
\begin{center}
\epsfysize=8cm \epsffile{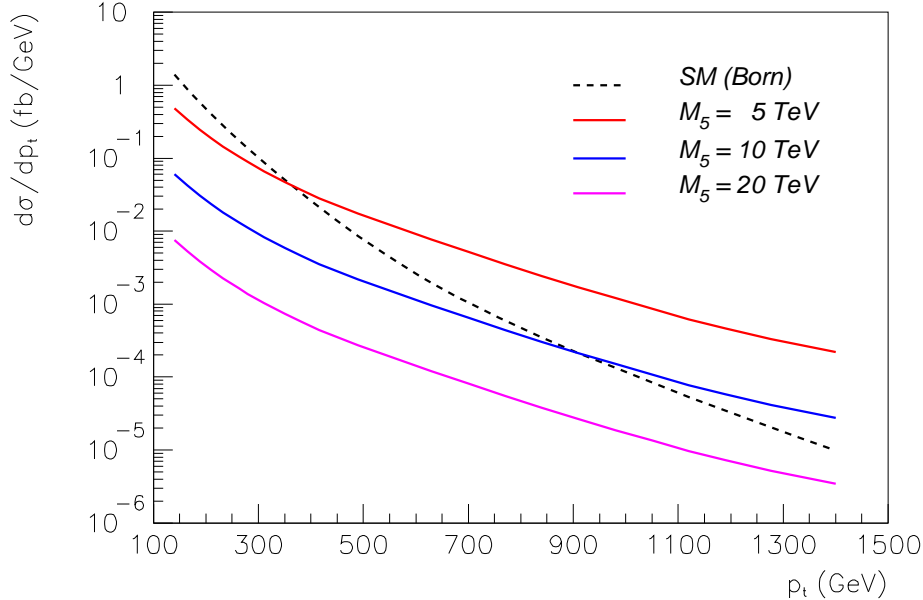}
\end{center}
\caption{The s-channel graviton contribution to the process $p \,
p \rightarrow \gamma \, \gamma + X$ for different values of
$\bar{M}_5$ (solid curves) vs. SM contribution (dashed curve) at
the LHC.} \label{fig:LHC_SM_grav}
\end{figure}

\begin{figure}[htb]
\begin{center}
\epsfysize=8cm \epsffile{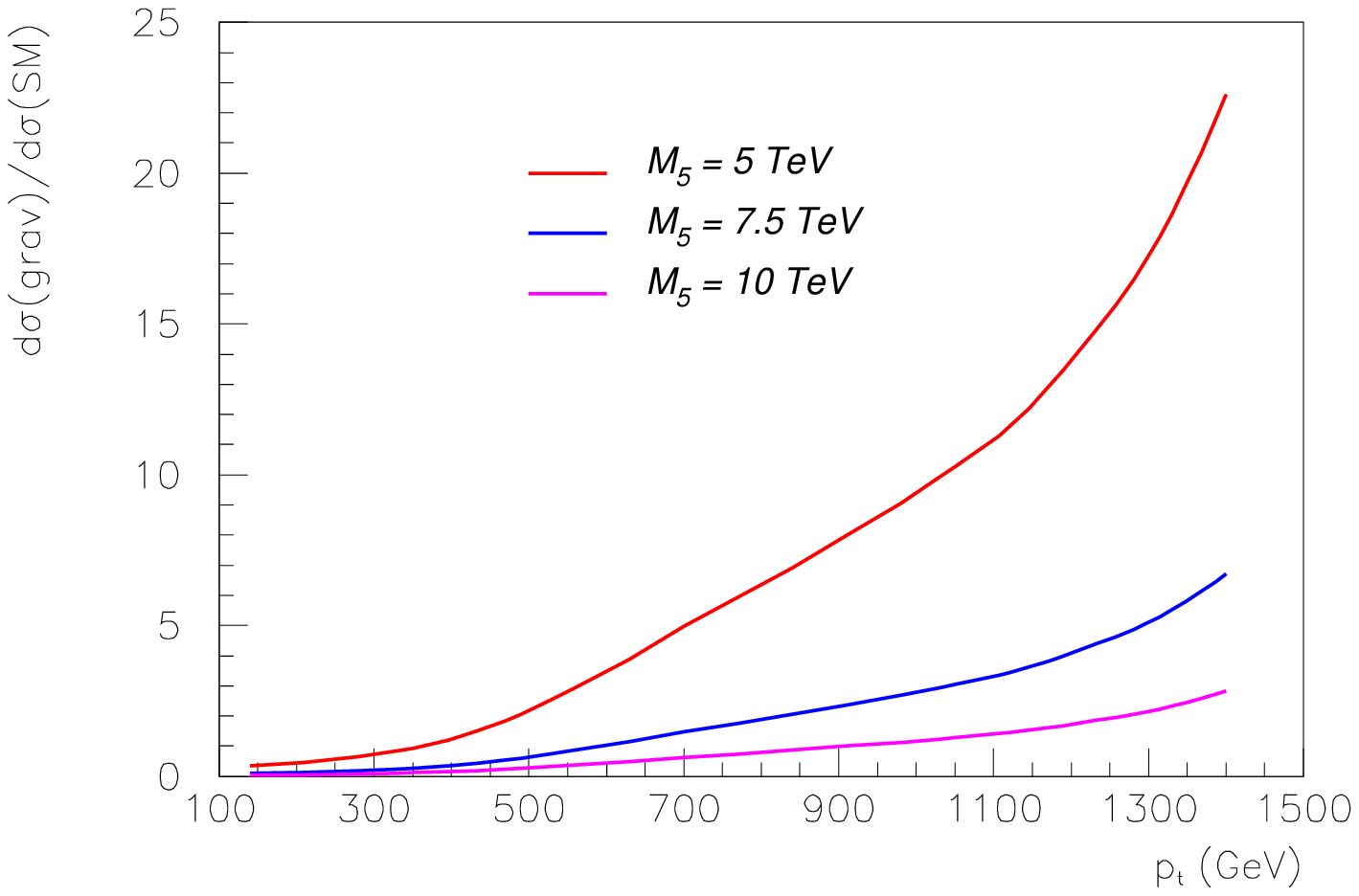}
\end{center}
\caption{The ratio of the gravity induced cross section to the SM
cross section for the diphoton production at the LHC as a function
of the photon transverse momentum.}
\label{fig:LHC_ratio}
\end{figure}

In order to obtain a correct search limit for the LHC, we have
calculated contributions of $s$-channel gravitons to
$p_{\perp}$-distributions of the final photons for different
values of $\bar{M}_5$ (see Fig.~\ref{fig:LHC_SM_grav}). The ratio
of the gravity induced term to the SM one is presented on the next
plot (see Fig.~\ref{fig:LHC_ratio}). The ratio rises monotonically
with $p_{\perp}$ for all $\bar{M}_5$. Its dependence on
$\bar{M}_5$ is in accordance with Eq.~\eqref{grav_vs_SM}. The
details of our calculations are discussed in
Section~\ref{sec:details}.

\begin{figure}[htb]
\begin{center}
\epsfysize=8cm \epsffile{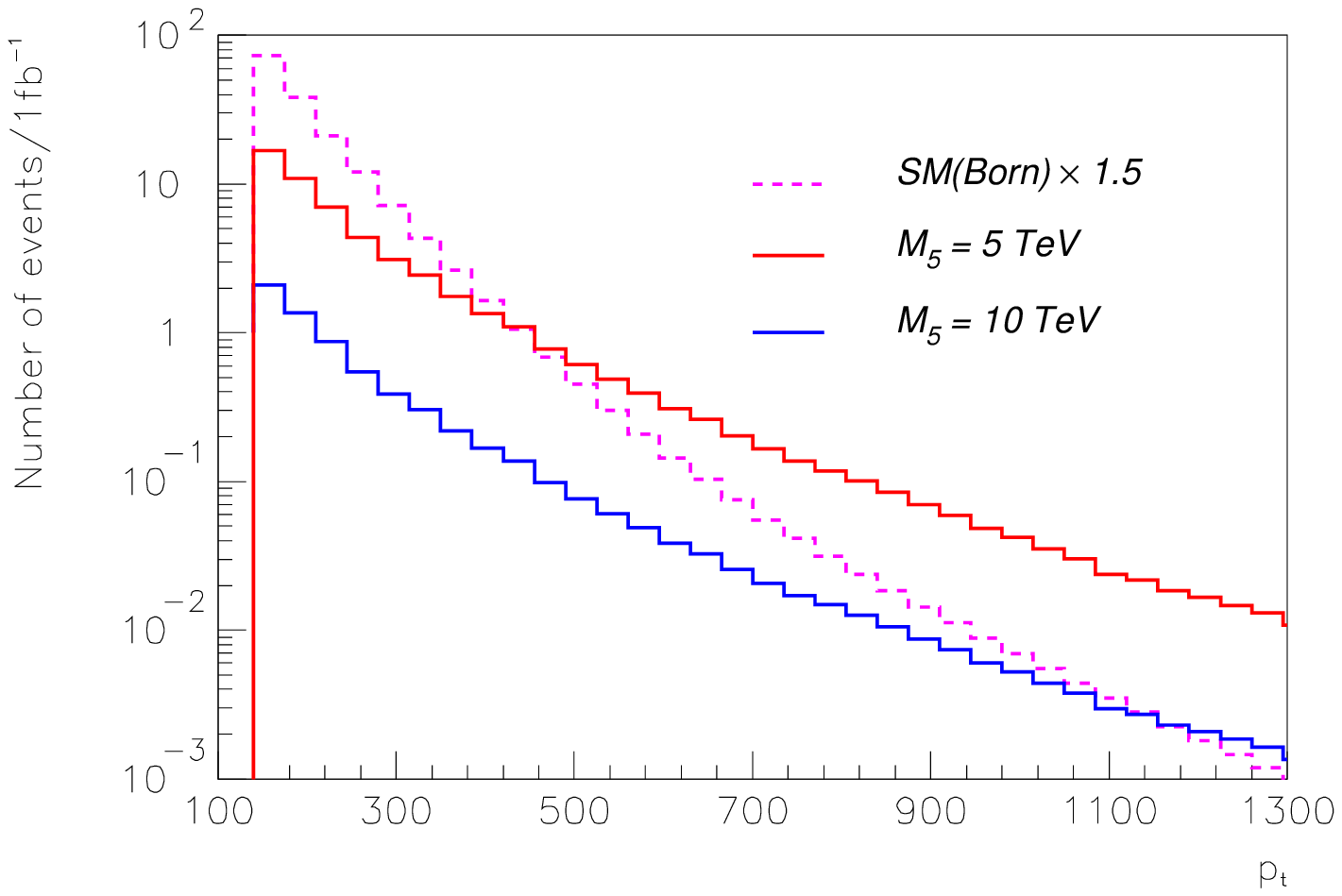}
\end{center}
\caption{The expected number of events per 35 GeV bin at the
integrated luminosity $\mathcal{L} = 1 \mathrm{\ fb}^{-1}$ for the
diphoton production at the LHC. The solid histograms denote the
gravity contributions, while the dashed one corresponds to the SM
(Born) term times K-factor 1.5.}
\label{fig:LHC_events}
\end{figure}

\begin{figure}[htb]
\begin{center}
\epsfysize=8cm \epsffile{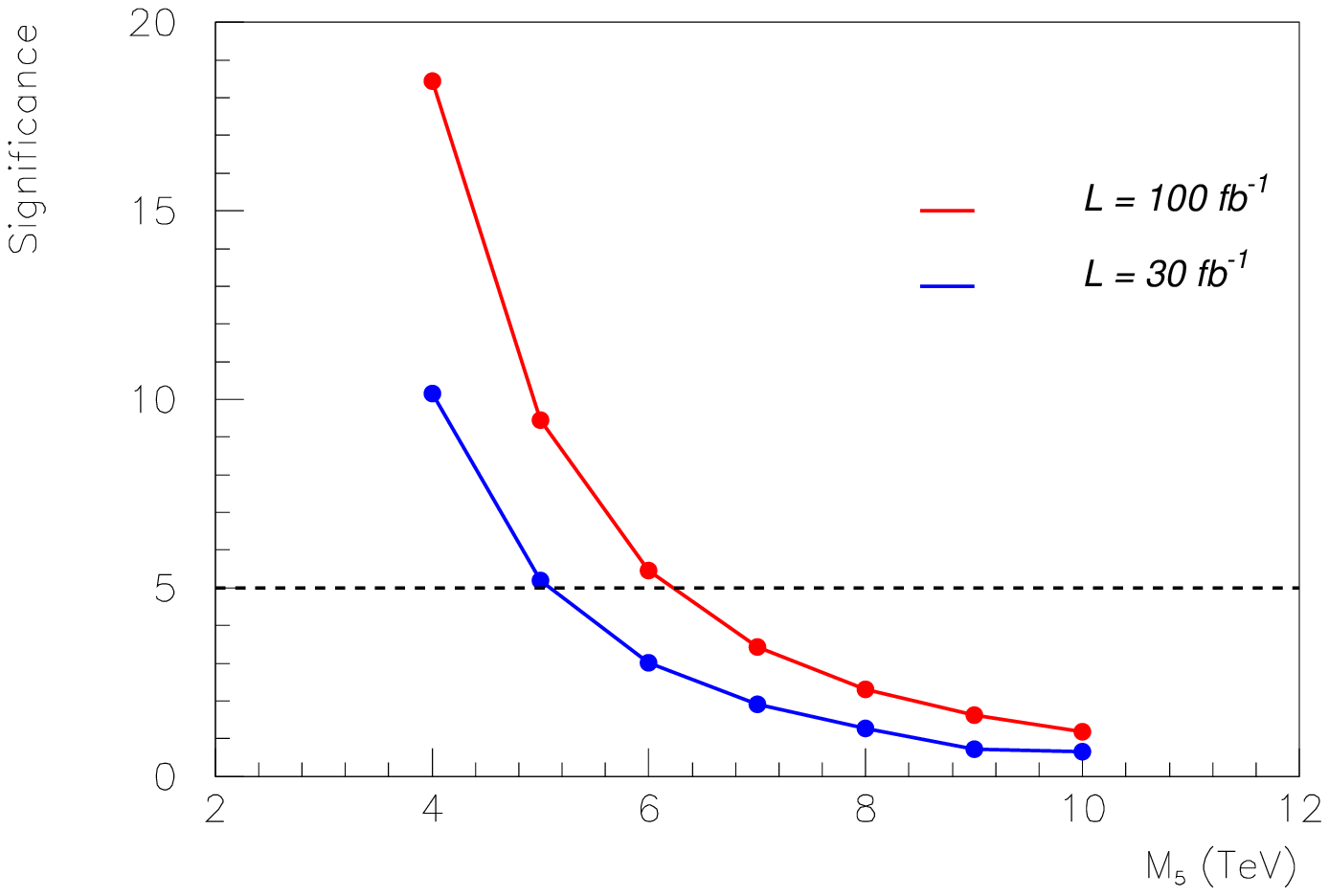}
\end{center}
\caption{The statistical significance for the process $p p
\rightarrow \gamma \gamma + X$ at the LHC as a function of the
5-dimensional (reduced) Planck scale.}
\label{fig:LHC_signif}
\end{figure}

Taking into accounts the K-factors described above, we obtain
histograms in Fig.~\ref{fig:LHC_events} which show a number of
events per 35 GeV bin at the integrated luminosity $\mathcal{L} =
1 \mathrm{\ fb}^{-1}$. A statistical significance as a function of
the 5-dimensional Planck scale is presented in
Fig.~\ref{fig:LHC_signif}. Thus, we obtain the discovery limit of
the LHC in the two-photon production in the RS1 model with the
small curvature:
\begin{equation}\label{search_limit_reduced}
\bar{M}_5 =
\left\{
\begin{array}{rl}
  6.3 \mathrm{\ TeV} \;, & \mathcal{L} = 100 \mathrm{\
fb}^{-1} \\
  5.1 \mathrm{\ TeV} \;,  & \mathcal{L} = 30 \mathrm{\
fb}^{-1} \\
\end{array}
\right.
\end{equation}
Let us stress that this limit \eqref{search_limit_reduced} do not
depend on the ratio $\kappa/\bar{M}_5$, contrary to the
conventional RS scenario~\eqref{metric_uncorr} in which both the
curvature $\kappa$ and $\bar{M}_5$ are of order of the
4-dimensional Planck mass \eqref{scale_relation_uncorr}.

In order to estimate systematic theoretical uncertainties, we have
calculated $p_{\perp}$-distributions for different values of the
PDF scale $\mu$: $\mu^2 = \hat{s}$, $\mu^2 = 2\hat{s}$, and $\mu^2
= \hat{s}/2$. It has appeared that an uncertainty in $d\sigma
(\mathrm{grav})$ related with the PDF scale varies from 1.3\% at
$p_{\perp} = 140$ GeV to
6.4\% at $p_{\perp} = 1$ TeV.%
\footnote{Note, it is lower values of $p_{\perp}$ that are
actually relevant for a calculation of the significance
$\mathbb{S}$.}
As for uncertainty in the ratio $d\sigma
(\mathrm{grav})/d\sigma(\mathrm{SM})$, it decreases
slowly from 2.8\% to 2.1\%, respectively.%
\footnote{All these numbers are \emph{insensitive} to the value of
$\bar{M}_5$.}
As a result, the systematic uncertainty for $\bar{M}_5$ from the
PDF scale amounts to $\Delta \bar{M}_5 = 17 $ GeV at $\mathcal{L}
= 30 \mathrm{\ fb}^{-1}$. Another systematic theoretical error
comes from an uncertainty in cross sections due to a certain
proton PDFs. Adopting that it does not exceed
2.7\%~\cite{Alekhin:05}, we get $\Delta \bar{M}_5 = 23$ GeV.

One of systematic experimental uncertainties contributing to the
number of events comes from luminosity measurements. The design
precision of the luminosity is 5\% at $\mathcal{L} = 1 \mathrm{\
fb}^{-1}$~\cite{CMS:TDR}. However, for measurements based on
$\mathcal{L} = 30 \mathrm{\ fb}^{-1}$ or more, it is assumed that
a 3\% uncertainty can be achieved~\cite{CMS:TDR} that results in
$\Delta \bar{M}_5 = 26$ GeV. The error in measurements of the
photon transverse momenta is less than 0.74\% at $p_{\perp} \geq
140$ GeV~\cite{CMS:TDR}.

Rather small values of $\Delta \bar{M}_5$ cited above may be
understood as follows. Consider, for instance, the case when
$N_{\mathrm{S}}$ and $N_{\mathrm{B}}$ increased by the factor $(1
+ \rho)$ due to an increase of the integrated luminosity ($\rho
\ll 1$). Then a corresponding variation of the significance
$\mathbb{S}$ can be compensated by much smaller variation of the
5-dimensional gravity scale: $\bar{M}_5 \rightarrow \bar{M}_5 +
\Delta \bar{M}_5 \simeq \bar{M}_5 \, (1 + \rho/6)$.

\begin{figure}[htb]
\begin{center}
\epsfysize=8cm \epsffile{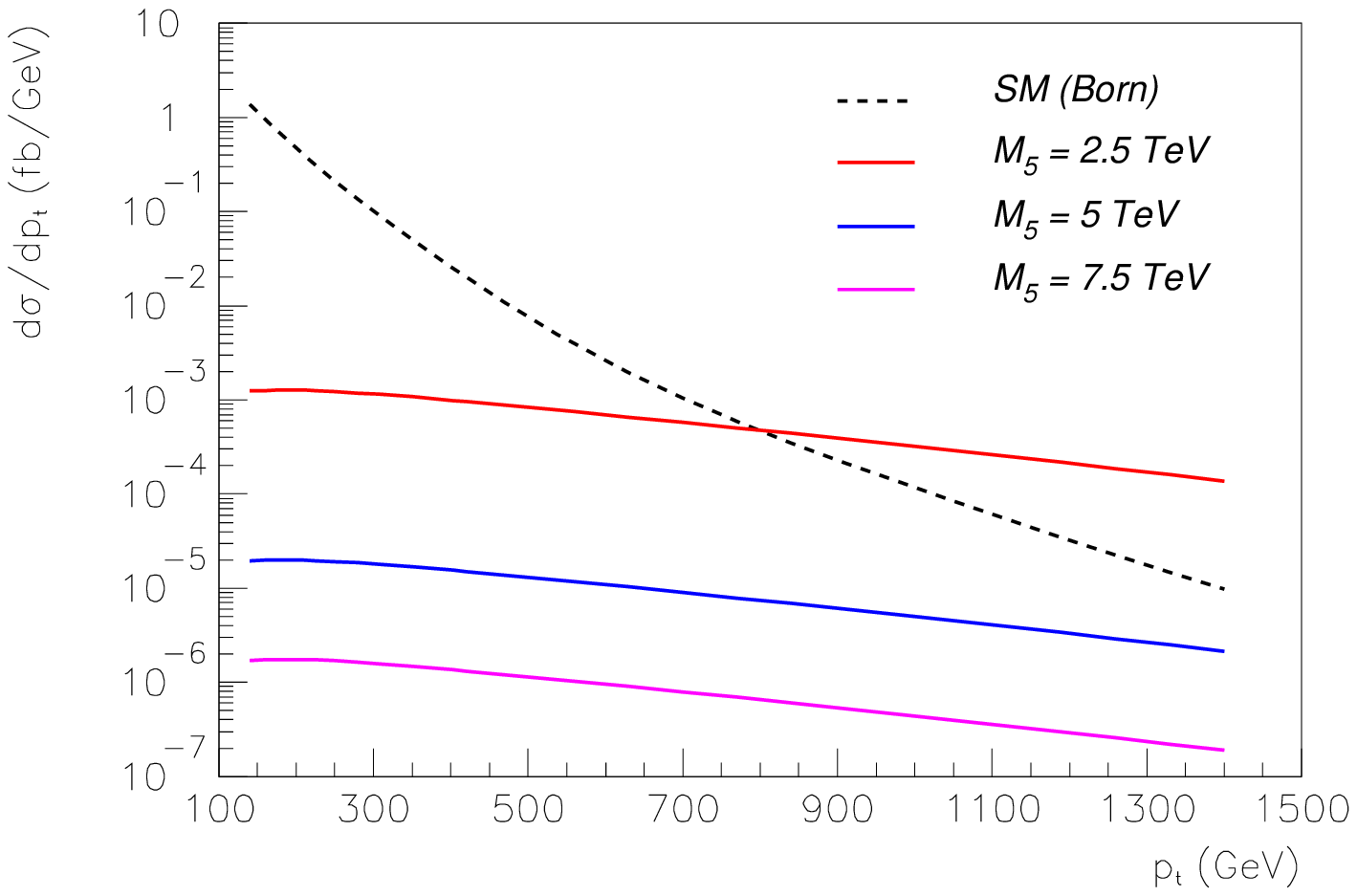}
\end{center}
\caption{The same as in Fig.~\ref{fig:LHC_SM_grav} (except for
values of $\bar{M}_5$), but calculations were done for \emph{zero
widths} of the KK gravitons (see formulae
\eqref{KK_sum_zero_widths}).}
\label{fig:LHC_SM_grav_zero}
\end{figure}

Finally, let us demonstrate that an ignorance of the graviton
widths would be a \emph{very rough approximation}.
Fig.~\ref{fig:LHC_SM_grav_zero} shows the gravity contribution to
the $p_{\perp}$-distribution calculated with the use of equations
\eqref{KK_sum_zero_widths}. The significant difference of
Fig.~\ref{fig:LHC_SM_grav_zero} from Fig.~\ref{fig:LHC_SM_grav}
can be explained as follows. As one can see from Appendix A, after
integrating over $s$, $ \mathrm{Re} \, \mathcal{S}(s)$ averages to
$0$, while $\mathrm{Im} \, \mathcal{S}(s)$ averages approximately
to 1. However, the expressions for the partonic cross
sections~\eqref{quark_cross_sec}, \eqref{gluon_cross_sec}, contain
quadratic term $|S(s)|^2$ as well as $s$-dependent factors.
Moreover, the region of integration over parton momentum fractions
$x_a$, $x_b$ (see Eq.~\eqref{cross_sec}) depends on s, since $x_a
x_b \geq 4p_{\perp}^2/s$.

Thus we conclude that the account of the width of the KK gravitons
is a crucial point for obtaining a correct result.

\section{Important features and details of calculations}
\label{sec:details}

In this Section, we will discuss some details of calculations of
the $p_{\perp}$-dist\-ri\-bution for the diphoton production at
the LHC, and analyze a possible dependence of this distribution on
the parameters $\bar{M}_5$ and $\kappa$. In particular, we will
show that  actually the distribution does not depend on $\kappa$
(or, equivalently, on the ratio $\kappa/\bar{M}_5$, which is taken
to to be small, see Eq.~\eqref{scale_relation}).

We start from formula~\eqref{cross_sec}. After changing variables,
\begin{equation}\label{tau}
x = x_a \;, \qquad \tau = x_a \, x_b \;,
\end{equation}
it has the following form:
\begin{align}\label{cross_sec_mod}
\frac{d \sigma}{d p_{\perp}^2}(p p \rightarrow \gamma \gamma + X)
&= \sum\limits_{a,b} \int\limits_{x_{\perp}^2}^1 \! \frac{d \tau
\sqrt{\tau}}{\sqrt{\tau - x_{\perp}^2}} \int\limits_{\tau}^1 \!
\frac{dx}{x} \, f_{a/p}(\mu^2, x) \, f_{b/p}(\mu^2, \tau/x)
\nonumber \\
&\times \frac{d \hat{\sigma}}{d\hat{t}}(a b \rightarrow \gamma
\gamma) \;,
\end{align}
where the partonic cross section $d \hat{\sigma}/d\hat{t}$ depends
on the function $\mathcal{S}(\hat{s})$~\eqref{grav_propagator}.
Since $\hat{s} = s \tau$, the latter is a function of $\tau$. It
is convenient to define
\begin{equation}\label{S_reduced}
\tilde{\mathcal{S}}(\tau) = \left[ -4\bar{M}_5^3 \sqrt{s \tau} \,
\right] \, \mathcal{S}(\tau) \;.
\end{equation}

Let $A = A_0 + a$, where $A_0 = (n_0 + 1/2) \pi$, $n_0$ being
\emph{an integer}, and $|a| \ll 1$. In a small vicinity of the
point $\tau_0 = (A_0 \kappa/\sqrt{s})$, the function
$\tilde{\mathcal{S}}(\tau)$ can be well approximated by the
espression
\begin{equation}\label{S_reduced_app}
\tilde{\mathcal{S}}(\tau) \simeq - \frac{2 \, a - \mathrm{2 \, i
\, \varepsilon_0}}{a^2 + \varepsilon_0^2} \;,
\end{equation}
where
\begin{equation}\label{epsilon_0}
\varepsilon_0  = \frac{\eta}{2} \Big( \frac{\sqrt{s
\tau_0}}{\bar{M}_5} \Big)^3 \;.
\end{equation}
We have taken into account that $\varepsilon_0 \ll 1$ for the
relevant values of $\sqrt{s}$ and $\bar{M}_5$ (i.e. for $\sqrt{s}
= 14$ TeV, $\bar{M}_5 \gtrsim 5$ TeV).

One can see from \eqref{S_reduced_app} that the real part,
$\mathrm{Re} \, \tilde{\mathcal{S}}(\tau)$, equals zero at $\tau =
\tau_0$ and has two extremal points at $\tau = \tau_0 \pm \delta$,
where%
\footnote{The relation $\tau = \tau_0 \pm \delta$ is equivalent to
the relation $a = \pm \, \varepsilon_0$.}
\begin{equation}\label{delta_tau}
\delta = \eta \, \tau_0^2 \, \frac{\kappa \, s}{\bar{M}_5^3} \;.
\end{equation}
As for the imaginary part, $\mathrm{Im} \,
\tilde{\mathcal{S}}(\tau)$, it has a very sharp peak at $\tau =
\tau_0$. Two peaks in the real part of the resonance are separated
by the distance $2\delta$, the width of the imaginary of the
resonance is also equal to $\Gamma_{\mathcal{S}} = 2\delta$.

\begin{figure}[htb]
\hskip 0cm \vbox to 5cm {\hbox to 7cm
{\epsfxsize=7cm\epsfysize=5cm\epsffile{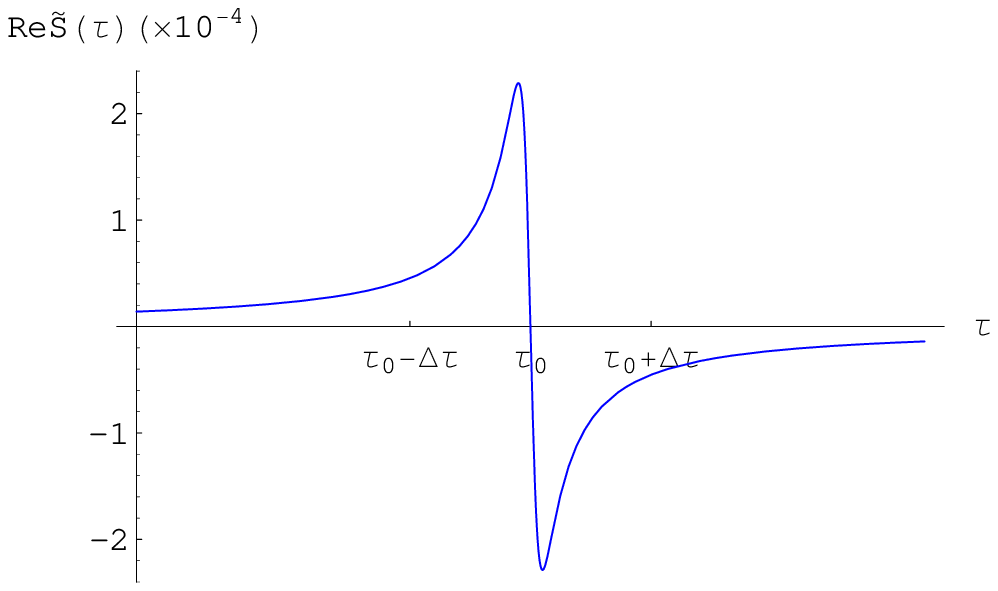}}} \hskip -0.6cm
\vbox to 5cm {\hbox to 7cm
{\epsfxsize=7cm\epsfysize=5cm\epsffile{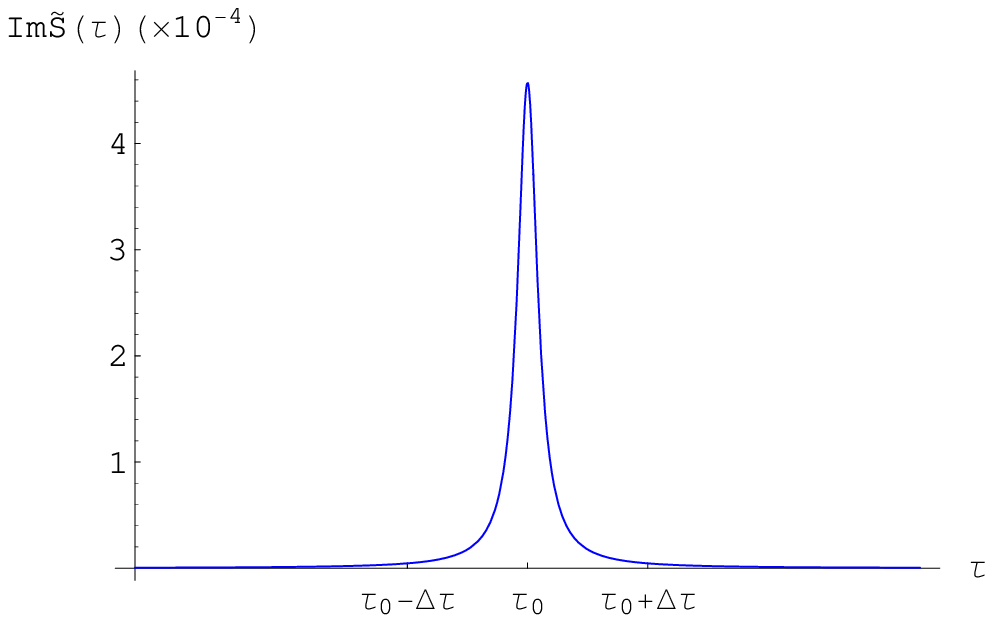}}} \caption{The
real and imaginary parts of one of the resonances in the function
$\tilde{\mathcal{S}}(\tau)$ which describes virtual graviton
contributions to partonic sub-processes. Both the curves
correspond to $\tau_0 = 0.02$, $\Delta \tau = 8.8 \cdot 10^{-9}$.}
\label{fig:S_1}
\end{figure}

\begin{figure}[htb]
\hskip 0cm \vbox to 5cm {\hbox to 7cm
{\epsfxsize=7cm\epsfysize=5cm\epsffile{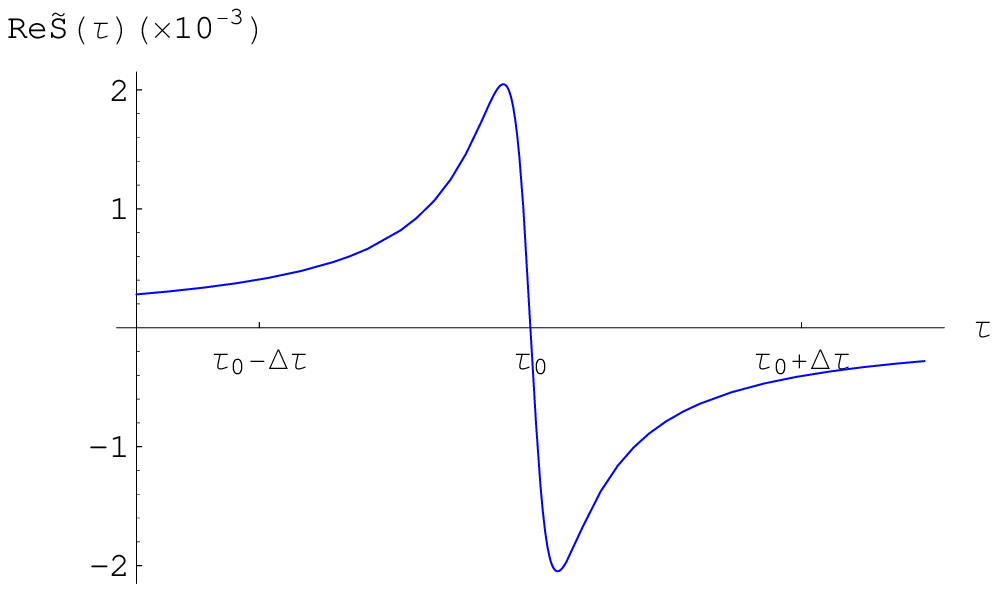}}} \hskip -0.6cm
\vbox to 5cm {\hbox to 7cm
{\epsfxsize=7cm\epsfysize=5cm\epsffile{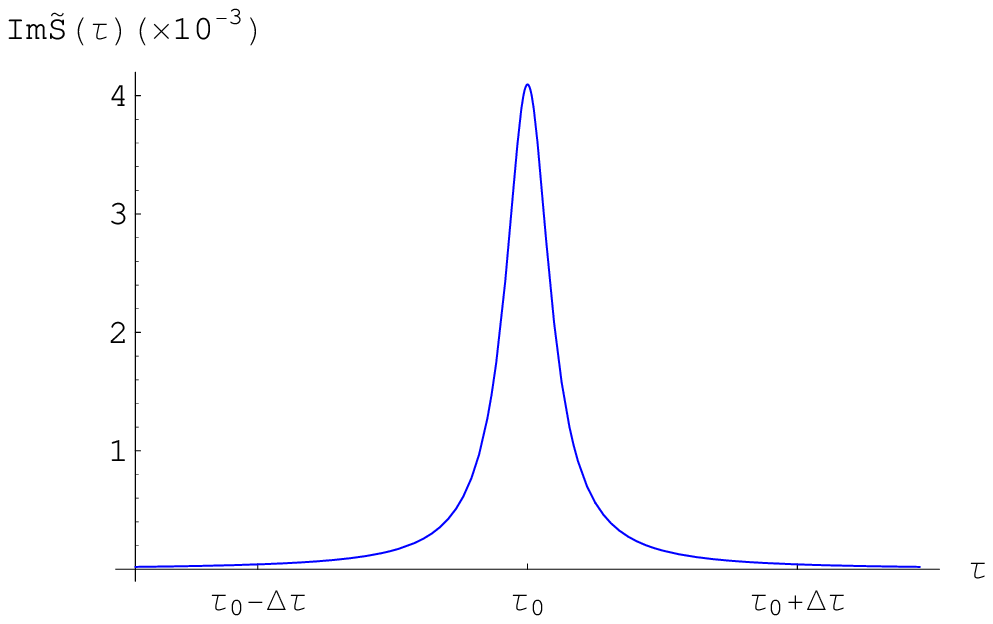}}} \caption{The
same as in Figs.~\ref{fig:S_1}, but for another peak in
$\tilde{\mathcal{S}}(\tau)$ located at $\tau_0 = 0.1$, with
$\Delta \tau = 2.2 \cdot 10^{-7}$.}
\label{fig:S_2}
\end{figure}

The forms of the real and imaginary parts of
$\tilde{\mathcal{S}}(\tau)$ are presented in Fig.~\ref{fig:S_1}
and Fig.~\ref{fig:S_2}, with $\Delta \tau = 10 \, \delta$ on both
the figures. All the curves were calculated using the values of
$\bar{M}_5 = 20$ TeV, $\kappa = 1$ GeV.

Let us describe Fig.~\ref{fig:S_1} first. The value of $\tau_0 =
0.02002$ was taken, that corresponds to the equation $\sqrt{s
\tau_0}/\kappa = (630 + 1/2) \pi$. Than we obtain that
$\Gamma_{\mathcal{S}} \simeq 1.8 \cdot 10^{-9}$. The calculations
show that the next resonance is located at the point $\tau_0 =
0.02008$. Thus, an average distance between neighboring peaks is
much larger than their widths. It reflects the fact that one
cannot regard a set of narrow resonances to be a continuous
spectrum~\cite{Kisselev:06}.

In the next two plots (see Fig.~\ref{fig:S_2}), $\tau_0 =
0.10004$, $n_0 = 1409$, and $\Gamma_{\mathcal{S}} \simeq 4.4 \cdot
10^{-8}$. A neighboring resonance with the number $n_0 = 1410$ is
located at $\tau_0 = 0.10018$.

The differential cross section of the process under consideration
is represented in the form:
\begin{equation}\label{cross_sec_sum}
d\sigma = d \sigma (\mathrm{SM}) + d\sigma (\mathrm{grav}) +
d\sigma (\mathrm{SM}\!-\!\mathrm{grav}) \;,
\end{equation}
where the last term comes from the interference between the SM and
graviton interactions. Since the SM amplitude is pure real, while
the real part of each graviton resonance is antisymmetric with
respect to its central point $\tau_0$ (see the first curves in
Figs.~\ref{fig:S_1}-\ref{fig:S_2}), the interference term in
\eqref{cross_sec_sum} has appeared to be negligible in comparison
with the pure gravity contribution (the second term in
\eqref{cross_sec_sum}) after integration in variable $\tau$.

The smaller is the value of $\tau_0$, the narrower is the peak and
larger is its height (compare, for instance, the curves in
Fig.~\ref{fig:S_1} and Fig.~\ref{fig:S_2}). The latter
circumstance means that only a smaller part of the graviton
resonances is significant for numerical calculations. The total
number of the graviton resonances which contribute to the
differential cross section is equal to $N = \mathrm{Int}[\sqrt{s}
\, (1 - x_{\perp})/(\kappa \pi)]$, with $\mathrm{Int}[z]$ being an
integer part of $z$.%
\footnote{The variable $x_{\perp}$ was defined
above~\eqref{x_trans}.}
Actually, the main contribution to $d\sigma (\mathrm{grav})$ comes
from, approximately, $(1/7)N$ first resonances, while the account
of the rest of $(6/7)N$ resonances results in a few percent
correction.

From all said above, we find that the gravity contribution to the
partonic cross section is proportional to
\begin{equation}\label{one_reson_contribution}
s^2 \left[ \frac{1}{\bar{M}_5^3 \sqrt{s} \, \varepsilon_0}
\right]^2 \! \delta \, N \sim \frac{1}{\bar{M}_5^3 \sqrt{s}} \;.
\end{equation}
In other words, we obtain that the differential cross section
\eqref{cross_sec_mod} \emph{does not depend on the curvature}, and
\begin{equation}\label{M5_dependence}
\frac{d\sigma (\mathrm{grav})}{d p_{\perp}} =
\frac{1}{\bar{M}_5^3} \, F(x_{\perp}, \ln s) \;,
\end{equation}
A weak $s$-dependence of $F(x_{\perp}, \ln s)$ comes from the PDF
scale. The numerical calculations \emph{do confirm} that the
differential cross section of the process \eqref{process} does not
depend on $\kappa$ and decreases as the third power of the
fundamental gravity scale $\bar{M}_5$.

At small values of $x_{\perp}$, the main contribution to the
$p_{\perp}$-distribution comes from the gluon-gluon
fusion~\eqref{gluon_fusion}. Assuming that $g(\mu^2, x) \sim
x^{-1}$, we get a \emph{rough} estimate from
\eqref{cross_sec_mod}:
\begin{equation}\label{p_trans_dependence}
\frac{d\sigma (\mathrm{grav})}{d p_{\perp}} \sim x_{\perp} \!\!
\int\limits_{x_{\perp}^2}^1 \! d\tau \, \frac{1}{\tau^{5/2} \,
\sqrt{\tau - x_{\perp}^2}} \, \ln\frac{1}{\tau} \sim
\frac{1}{p_{\perp}^3} \ln\frac{1}{p_{\perp}} \;.
\end{equation}
This form of $d\sigma (\mathrm{grav})/d p_{\perp}$ is in
satisfactory agreement with our numerical calculations at
$p_{\perp} \ll \sqrt{s}$ (see Figs.~\ref{fig:Tevatron_SM_grav},
\ref{fig:LHC_SM_grav}).

In order to get a correct numerical result, we divided a region of
integration in variable $\tau$ into small subregions around the
resonance peaks. Since a typical value of $N/7$ exceeds $600$ ($N
= 4367$ for $x_{\perp} = 0.02$, $\kappa = 1$ GeV), it took lots of
computer time. Fortunately, a number of integrations could be
reduced if we use the facts that the gravity cross section is
actually independent of $\kappa$, while $N \sim \kappa^{-1}$.



\section{Conclusions}

In the present paper the model with one large warped extra
dimension (i.e. the RS1 scheme with the small curvature $\kappa$)
is studied~\cite{Kisselev:05,Kisselev:06}. In such an approach,
the reduced fundamental gravity scale $\bar{M}_5$ lies in the TeV
region (i.e. varies from one to tens TeV), and $\kappa \ll
\bar{M}_5$. The mass spectrum is similar to that in the ADD
model~\cite{Arkani-Hamed:98} with one flat extra dimension.

We have calculated the $p_{\perp}$-distribution for the process $p
p \rightarrow \gamma \gamma + X$ at the LHC, with $p_{\perp}$
being the transverse momenta of the final photons. The LHC
discovery limit on the reduced fundamental gravity scale
$\bar{M}_5$  has been obtained which is given by
Eq.~\eqref{search_limit_reduced}. Remembering relation between the
5-dimensional Planck mass $M_5$ and its reduced value
$\bar{M}_5$~\eqref{reduced_scale}, we find the reach of the LHC in
the search for the RS gravitons decaying into diphoton channel:
\begin{equation}\label{search_limit}
M_5 = \left\{
\begin{array}{rl}
  11.6 \mathrm{\ TeV} \;, & \mathcal{L} = 100 \mathrm{\
fb}^{-1} \\
  9.4 \mathrm{\ TeV} \;,  & \mathcal{L} = 30 \mathrm{\
fb}^{-1} \\
\end{array}
\right.
\end{equation}

In the conventional RS scenario~\cite{Randall:99}, both $\kappa$
and $\bar{M}_5$ are of order of the 4-dimensional Planck mass,
$\kappa \sim \bar{M}_5 \sim M_{\mathrm{Pl}}$. A search limit on
the lightest graviton mass depends crucially on the ratio
$\kappa/\bar{M}_5$. On the contrary, our bounds
\eqref{search_limit} do not depend on $\kappa$, since the gravity
cross sections are insensitive to its value (provided $\kappa \ll
M_5$).

We have shown that neglecting the width of the KK gravitons would
give us incorrect results. A zero width approximation is valid
only if an effective collision energy of partonic sub-processes is
at least 3.5 times larger than $\bar{M}_5$ (see
Eq.~\eqref{sum_zero_widths_app}).



\section*{Acknowledgements}

I am grateful to V.A. Petrov for fruitful discussions and critical
remarks.





\setcounter{equation}{0}
\renewcommand{\theequation}{A.\arabic{equation}}

\section*{Appendix A}
\label{app:A}

In this Appendix we will present a result of computations of the
function $\mathcal{S}(s)$ with the use of three different
equations \eqref{grav_propagator}, \eqref{KK_sum_zero_widths}, and
\eqref{KK_sum}.

\begin{figure}[htb]
\begin{center}
\epsfysize=6cm \epsffile{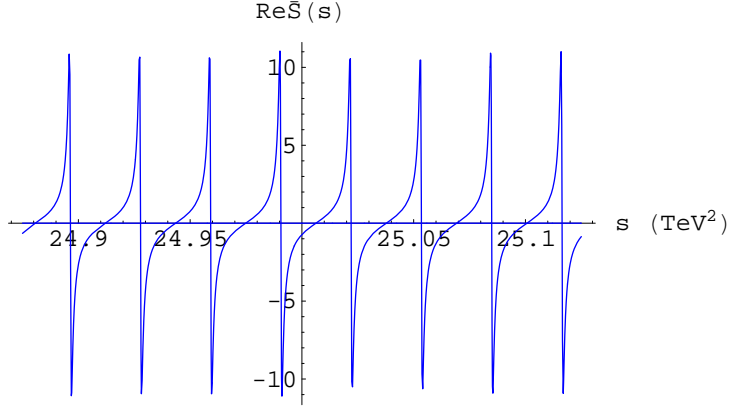}
\end{center}
\caption{The real part of $\mathcal{\bar{S}}(s)$ calculated by
using Eq.~\eqref{grav_propagator}. The line
$\mathrm{Re}\,\mathcal{\bar{S}}(s) = 0$ corresponds to
Eq.~\eqref{KK_sum_zero_widths}. The values of the parameters are:
$\bar{M}_5 = 5$ TeV, $\kappa = 1$ GeV.}
\label{fig:ReS_full}
\end{figure}

\begin{figure}[htb]
\begin{center}
\epsfysize=6cm \epsffile{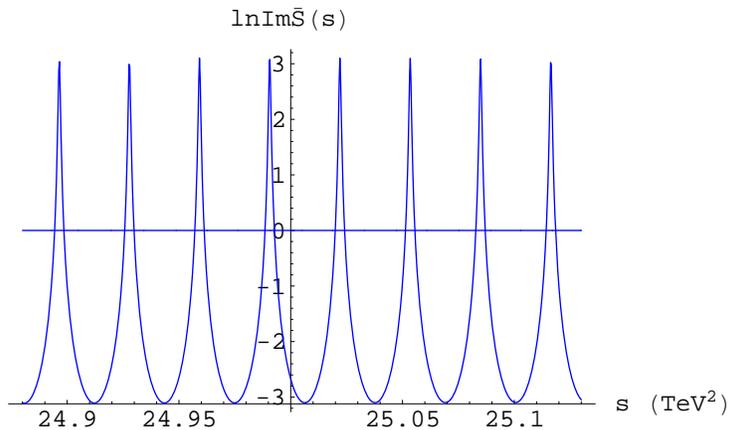}
\end{center}
\caption{The function $\ln \mathrm{Im}\mathcal{\bar{S}}(s)$
calculated by using Eq.~\eqref{grav_propagator}. The horizontal
line corresponds to $\mathrm{Im}\mathcal{\bar{S}}(s) = 1$ (see
Eq.~\eqref{KK_sum_zero_widths}). The values of the parameters are
the sane as in Fig.~\ref{fig:ReS_full}.}
\label{fig:ImS_full}
\end{figure}

Let us define the dimensionless function:
\begin{equation}\label{S_modified}
\bar{\mathcal{S}}(s) = \left[ -2\bar{M}_5^3 \sqrt{s} \, \right] \,
\mathcal{S}(s) \;.
\end{equation}
The real and imaginary parts of $\mathcal{\bar{S}}(s)$ in the
energy region around the point $\sqrt{s_0} = 5$ TeV are shown in
Fig.~\ref{fig:ReS_full} and Fig.~\ref{fig:ImS_full}, respectively.
The values of the parameters were chosen to be $\bar{M}_5 = 5$
TeV, $\kappa = 1$ GeV. The resonance structures on both figures
are obtained by using formula \eqref{grav_propagator}, while the
solid lines correspond to ``zero width'' equation
\eqref{KK_sum_zero_widths}. One can see that formula
\eqref{KK_sum_zero_widths} is incorrect.

\begin{figure}[htb]
\begin{center}
\epsfysize=6cm \epsffile{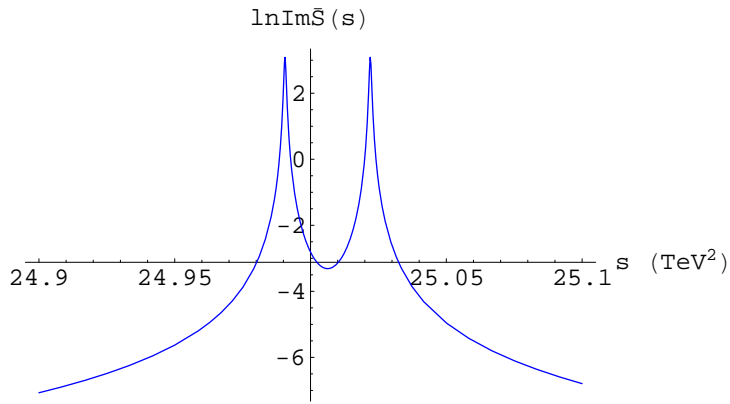}
\end{center}
\caption{The imaginary part of $\mathcal{\bar{S}}(s)$ calculated
by using Eq.~\eqref{KK_sum}. Only two terms in the sum are taken
into account. The values of the parameters are the same as in
Fig.~\ref{fig:ImS_full}.}
\label{fig:ImS_2_peaks}
\end{figure}

To demonstrate that our formula \eqref{grav_propagator} is a
correct expression for $\mathcal{S}(s)$, one should calculate the
sum \eqref{KK_sum}. It appeared that only the terms closed to
$n=n_0$, where $m_{n_0} = x_{n_0} \kappa \simeq \sqrt{s_0}$, are
important in the sum. For a sake of simplicity, we consider only
the imaginary part of $\mathcal{\bar{S}}(s)$. In
Fig.~\ref{fig:ImS_2_peaks} we present the function $\ln
\mathrm{Im}\mathcal{\bar{S}}(s)$ calculated for the case when only
two terms in the sum, namely, $n=n_0$ and $n=n_0 + 1$, are taken
into account. The curve in the next figure corresponds to the case
when already eight neighboring terms are taken into account.%
\footnote{For a wider region of $s$ more terms have to be
considered in the sum~\eqref{KK_sum}.}
%
\begin{figure}[htb]
\begin{center}
\epsfysize=6cm \epsffile{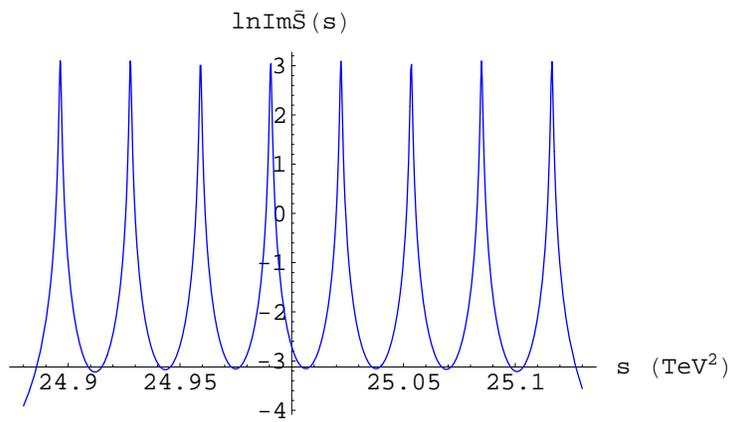}
\end{center}
\caption{The same as in Fig.~\ref{fig:ImS_2_peaks} but with eight
terms taken into account in Eq.~\eqref{KK_sum}.}
\label{fig:ImS_8_peaks}
\end{figure}

A comparison of Fig.~\ref{fig:ImS_8_peaks} with
Fig.~\ref{fig:ImS_full} demonstrates us that our formula
\eqref{grav_propagator} is a nice approximation of the original
expression~\eqref{KK_sum}.



\begin{thebibliography}{99}

\bibitem{Randall:99}
L. Randall and R. Sundrum, Phys. Rev. Lett.  {\bf 83}, 3370
(1999).
\bibitem{Boos:02}
E.E.~Boos, Yu.A.~Kubyshin,, M.N.~Smolyakov and I.P.~Volobuev,
Class. Quan. Grav. {\bf 19}, 4591 (2002); E.E.~Boos,
Yu.S.~Mikhailov, M.N.~Smolyakov and I.P.~Volobuev, Nucl. Phys. B
{\bf 717}, 19 (2005).
\bibitem{Giudice:05}
G.F. Giudice, T. Plehn and A. Strumia, Nucl. Phys. B {\bf 706},
455 (2005).
\bibitem{Kisselev:05}
A.V. Kisselev and V.A. Petrov, Phys. Rev. D {\bf 71}, 124032
(2005).
\bibitem{Kisselev:06}
A.V. Kisselev, Phys. Rev. D {\bf 73}, 024007 (2006). .
\bibitem{Kisselev:diff}
A.V. Kisselev, V.A. Petrov and R.A. Ryutin, Phys. Lett.B {\bf
630}, 100 (2005).
\bibitem{Kisselev:ILC}
A.V. Kisselev, JHEP 03, 006 (2007)
\bibitem{Arkani-Hamed:98}
N. Arkani-Hamed, S. Dimopoulos and G. Dvali, Phys. Lett. B {\bf
429}, 263 (1998); I.~Antoniadis, N.~Arkani-Hamed, S.~Dimopoulos
and G. Dvali, Phys. Lett. B {\bf 436}, 257 (1998);
N.~Arkani-Hamed, S.~Dimopoulos and G. Dvali, Phys. Rev. D {\bf
59}, 086004 (1999).
\bibitem{CMS:TDR}
CMS Collaboration 2006 Technical Design Report, Volume I,
CERN/LHCC 2006-001, CMS TDR 8.1; CMS Collaboration 2007 Technical
Design Report, Volume II: Physics Perfomance, J. Phys. G: Nucl.
Part.Phys. {\bf 34}, 995 (2007).
\bibitem{Kisselev:05_2}
A.V. Kisselev, Eur. Phys. J. C {\bf 42}, 217 (2005).
\bibitem{LEP_limit}
K. Klein, Proceeding of the 34th International Conference on High
Energy Physics (ICHEP 2006), July 26 - August 2, 2006, Moscow,
Russia, p.~1133; S.~Ask, V.~Hedberg and F.-L.~Navarria,
Contributed paper for the 23rd International Symposium on
Lepton-Photon Interactions at High Energy (LP07), August 13-18,
2007, Daegu, Korea.
\bibitem{Tevatron_ADD}
B. Abbott \emph{et al.}, (D\O~Collaboration), Phys Rev. Lett. {\bf
86}, 1156 (2001); D\O~Note 4336-Conf (2004); D. Acosta \emph{et
al.} (CDF Collaboration), Phys Rev. Lett. {\bf 95}, 022003 (2005).
\bibitem{Tevatron_RS}
V.M. Abazov \emph{et al.}, (D\O~Collaboration), Phys Rev. Lett.
{\bf 95}, 091801 (2005); Phys Rev. Lett., accepted for publication
(arXiv:07103338 (2007)).
\bibitem{CDF_preliminary}
A. Pronko, Proceeding of the 34th International Conference on High
Energy Physics (ICHEP 2006), July 26 - August 2, 2006, Moscow,
Russia, p.~1163.
\bibitem{Alekhin:05}
S.I. Alekhin, JETP Lett. {\bf 82}, 628 (2005).
\end{thebibliography}
\end{document}